\documentstyle[12pt,graphicx]{article}                
\def\Journal#1#2#3#4{{#1} {\bf #2}, #3 (#4)}

\def\NPB{{\em Nucl. Phys.} B}
\def\PLB{{\em Phys. Lett.}  B}
\def\PRL{\em Phys. Rev. Lett.}
\def\PRD{{\em Phys. Rev.} D}

\begin{document}
%This is dvips(k) 5.86 Cobegin{document}
\baselineskip 18pt
%t
\def\today{\ifcase\month\or
 January\or February\or March\or April\or May\or June\or
 July\or August\or September\or October\or November\or December\fi
 \space\number\day, \number\year}
\def\thebibliography#1{\section*{References\markboth
 {References}{References}}\list
 {[\arabic{enumi}]}{\settowidth\labelwidth{[#1]}
 \leftmargin\labelwidth
 \advance\leftmargin\labelsep
 \usecounter{enumi}}
 \def\newblock{\hskip .11em plus .33em minus .07em}
 \sloppy
 \sfcode`\.=1000\relax}
\let\endthebibliography=\endlist
\def\lsim{\ ^<\llap{$_\sim$}\ }
\def\gsim{\ ^>\llap{$_\sim$}\ }
\def\r2{\sqrt 2}
\def\beq{\begin{equation}}
\def\eeq{\end{equation}}
\def\beqn{\begin{eqnarray}}
\def\eeqn{\end{eqnarray}}
\def\rmuu{\gamma^{\mu}}
\def\rmud{\gamma_{\mu}}
\def\PL{{1-\gamma_5\over 2}}
\def\PR{{1+\gamma_5\over 2}}
\def\sinW2{\sin^2\theta_W}
\def\AEM{\alpha_{EM}}
\def\mul{M_{\tilde{u} L}^2}
\def\mur{M_{\tilde{u} R}^2}
\def\mdl{M_{\tilde{d} L}^2}
\def\mdr{M_{\tilde{d} R}^2}
\def\mz2{M_{z}^2}
\def\c2b{\cos 2\beta}
\def\au{A_u}         
\def\ad{A_d}
\def\cob{\cot \beta}
\def\v#1{v_#1}
\def\tb{\tan\beta}
\def\epem{$e^+e^-$}
\def\KK{$K^0$-$\bar{K^0}$}
\def\wi{\omega_i}
\def\xj{\chi_j}
\def\Wmu{W_\mu}
\def\Wnu{W_\nu}
\def\m#1{{\tilde m}_#1}
\def\mH{m_H}
\def\mw#1{{\tilde m}_{\omega #1}}
\def\mx#1{{\tilde m}_{\chi^{0}_#1}}
\def\mc#1{{\tilde m}_{\chi^{+}_#1}}
\def\mwi{{\tilde m}_{\omega i}}
\def\mxi{{\tilde m}_{\chi^{0}_i}}
\def\mci{{\tilde m}_{\chi^{+}_i}}
\def\mz{M_z}
\def\sw{\sin\theta_W}
\def\cw{\cos\theta_W}
\def\cb{\cos\beta}
\def\sb{\sin\beta}
\def\rwi{r_{\omega i}}
\def\rxj{r_{\chi j}}
\def\rfp{r_f'}
\def\Kik{K_{ik}}
\def\Fq2{F_{2}(q^2)}
\def\f{\({\cal F}\)}
\def\d1{{\f(\tilde c;\tilde s;\tilde W)+ \f(\tilde c;\tilde \mu;\tilde W)}}
%%%%%%%%%%%%%%%%%%%%%%%%%%%%%%%%%%
\def\tw{\tan\theta_W}
\def\sec2w{sec^2\theta_W}
\def\m12{m_{\frac{1}{2}}}
\def\12{$\frac{1}{2}$}
\def\m0{$m_0$}
%%%%%%%%%%%%%%%%%%%%%%%%%%%%%%%%%%

\begin{titlepage}

\begin{center}
{\large {\bf  Weak Isospin Violations in Charged and Neutral Higgs Couplings
from  SUSY Loop Corrections }}\\
\vskip 0.5 true cm
\vspace{2cm}
\renewcommand{\thefootnote}
{\fnsymbol{footnote}}
 Tarek Ibrahim$^{a,b}$ and Pran Nath$^{b}$  
\vskip 0.5 true cm
\end{center}

\noindent
{a. Department of  Physics, Faculty of Science,
University of Alexandria,}\\
{ Alexandria, Egypt\footnote{: Permanent address of T.I.}}\\ 
{b. Department of Physics, Northeastern University,
Boston, MA 02115-5000, USA } \\
%\footnote{ $\dagger$ : Permanent address}
\vskip 1.0 true cm
\centerline{\bf Abstract}
\medskip
Supersymmetric QCD and supersymmetric electroweak loop corrections to the  
violations of weak isospin to Yukawa couplings are  investigated.
Specifically it involves  an analysis of the supersymmetric loop corrections
to the Higgs couplings to the third generation quarks and
leptons. Here we analyze the SUSY loop corrections to the charged
Higgs couplings which are then compared with the supersymmetric loop
corrections to the neutral Higgs couplings previously 
computed. It is found that the weak 
isospin violations can be quite significant, i.e, as much as
 40-50\%  or more of the total loop correction to the Yukawa coupling. 
  The effects  of CP phases are also studied and it is found that
 these effects can either enhance or suppress the weak isospin 
 violations. 
  We also investigate the weak isospin violation effects on the branching ratio
 $BR(H^-\rightarrow  \bar t b)/ 
 BR(H^-\rightarrow  \bar \nu_{\tau}\tau^-)$ and show that the effects 
 are sensitive to CP phases. Thus an accurate
 measurement of this  branching ratio along with the branching ratio
 of the neutral Higgs boson decays
  %$BR(h^0\rightarrow \bar b b)/BR(h^0\rightarrow \bar \tau \tau)$
  can provide a measure of weak isospin violation along with providing
 a clue to the presence of supersymmetry.
 \end{titlepage}

\section{Introduction}
In this paper we investigate the effects of supersymmetric QCD and
supersymmetric electroweak corrections to the violations of weak isopsin
in the couplings of Higgs to quarks and leptons\cite{carena2002}. 
Specifically, we compute in this paper the
supersymmetric loop effects to the couplings of the charged Higgs
with quarks and leptons. These are then compared with the 
supersymmetric corrections to the couplings of the neutral Higgs.
 We also study the effects of CP phases on the charged Higgs 
 couplings. The CP phases that
 appear in the soft parameters, however, are subject to strong
 constraints from the EDMs of the electron\cite{eedm}, and of 
  the neutron\cite{nedm} and of
 the $Hg^{199}$ atom\cite{atomic}. 
 The mechanisms available for the suppression of the EDMs  
 associated with large phases consist of the cancellation 
 mechanism\cite{incancel,olive}, the mass suppression\cite{na} 
 and phases just in the third 
 generation\cite{chang} among others\cite{Ibrahim:2002ry,yamaguchi}.
 Large CP phases affect a variety of low energy phenomena including the
 Higgs sector. One such phenomenon is the CP even -CP odd mixing 
 among the neutral Higgs bosons which has been studied in great 
 detail\cite{pilaftsis,inhiggs,Carena:2001fw}. 
 Here we study the effects of CP phases on the charged 
 Higgs couplings to quarks and leptons and also the  effect of
 CP phases on the breakdown of the weak isospin invariance. 
 The outline of the rest of the paper is as follows: In Sec.2 
 we give the basic formalism. In Sec.3 we compute the SUSY QCD
 and SUSY electroweak loop correction to the charged Higgs couplings 
 to third generation
 quarks and leptons.  In Sec.4 we
 analyze the loop corrections to the charged Higgs decays 
  $H^-\rightarrow  \bar t b$ and 
 $H^-\rightarrow  \bar \nu_{\tau}\tau^-$. In Sec.5 we give a
 numerical analysis of the loop effects and estimate the 
 sizes of the loop corrections to violations of weak isospin.
 We also study the effects of CP phases on the charged Higgs couplings
 and on the ratio of the  decay branching ratios of the charged Higgs
 to $\bar t b$ and   $\bar \nu_{\tau}\tau^-$, Finally, the 
 conclusions are given in Sec.6. 
\section{The basic formalism}
We will use the framework of MSSM 
which contains two doublets of Higgs and for the soft breaking
sector we will use the extended sugra framework\cite{msugra}
 with nonuniversalities and with CP phases. Thus for the
 Higgs sector we have
\beqn
(H_1)= \left(\matrix{H_1^1\cr
 H_{1}^2}\right),~~
(H_2)= \left(\matrix{H_{2}^1\cr
             H_2^2}\right)
\label{a}
\eeqn
%\beqn
%(H_1)
% =\frac{1}{\sqrt 2} 
%\left(\matrix{v_1+\phi_1+i\psi_1\cr
%             H_1^-}\right),~~~
%(H_2) 
%=\frac{e^{i\theta_H}}{\sqrt 2} \left(\matrix{H_2^+ \cr
%             v_2+\phi_2+i\psi_2}\right)
%\label{2c}
%\eeqn
The components of $H_1$ and $H_2$ interact with the quarks and 
the leptons at the tree level through\cite{gunion} 
\beqn\label{b}
-{\cal {L}}=  \sum_{f=b,\tau} h_f \bar f_R f_L H_1^1 
+h_t \bar t_R t_L H_2^2 -h_b \bar b_R t_L H_1^2 
-h_t \bar t_R b_L H_2^{1} -h_{\tau}\bar \tau_R \nu_L H_1^2
+ H.c.
\eeqn
The loop corrections produce shifts in these couplings and generate
new ones as follows 
\beqn\label{c}
-{\cal {L}}_{eff}=\sum_{f=b,\tau} (h_f +\delta h_f)\bar f_R f_L H_1^1
+\Delta h_f \bar f_R f_L H_2^{2*} \nonumber\\
+(h_t+\delta h_t) \bar t_R t_L H_2^2  + \Delta h_t \bar t_R t_L H_1^{1*} 
-(h_b+\overline{\delta h_b}) \bar b_R t_L H_1^2  + 
\overline{\Delta h_b} \bar b_R t_L H_2^{1*}\nonumber\\
-(h_t+\overline{\delta h_t}) \bar t_R b_L H_2^1  + 
\overline{\Delta h_t} \bar t_R b_L H_1^{2*}
 -(h_{\tau}+\overline{\delta h_{\tau})} \bar \tau_R \nu_L H_1^2  + 
\overline{\Delta h_{\tau}} \bar \tau_R \nu_L H_2^{1*} +H.c.
\eeqn
where "*" has been used to get a gauge invariant ${\cal{L}}_{eff}$.
We rewrite Eq.~(\ref{c}) in a form which is  illustrative

\beqn\label{c1}
-{\cal {L}}_{eff}=\epsilon_{ij} [(h_b+\delta h_b)\bar b_R H_1^iQ_L^j
+ (h_{\tau}+\delta h_{\tau})\bar \tau_R H_1^il_L^j
+ (h_t+\delta h_t)\bar t_R Q_L^iH_2^j ] \nonumber\\
+[\Delta h_b \bar b_R Q_L^iH_2^{i*} + \Delta h_{\tau} 
\bar \tau_R l_L^iH_2^{i*} 
+\Delta h_t \bar t_R Q_L^iH_1^{i*}]
-{\cal { L}}_{violation}+H.c.
\eeqn
where
\beqn\label{c2}
-{\cal {L}}_{violation}=
\{- (\overline{\delta h_b} -\delta h_b) \bar b_R t_L H_1^2
+ (\overline{\Delta h_b} -\Delta h_b) \bar b_R t_L H_2^{1*}\nonumber\\
- (\overline{\delta h_{\tau}} -\delta h_{\tau}) \bar \tau_R \nu_L H_1^2
+ (\overline{\Delta h_{\tau}} -\Delta h_{\tau}) \bar \tau_R \nu_L H_2^{1*}
\nonumber\\
- (\overline{\delta h_t} -\delta h_t) \bar t_R b_L H_2^1
+ (\overline{\Delta h_t} -\Delta h_t) \bar t_R b_L H_1^{2*}\}
\eeqn
The corrections $\delta h_b$, $\delta h_{\tau}$, $\delta h_t$,
$\Delta h_b$, $\Delta h_{\tau}$,and  $\Delta h_t$ have 
been calculated in Ref.\cite{Ibrahim:2003ca,Ibrahim:2003cv,Ibrahim:2003jm} 
with the inclusion of CP phases.
Their effects on the decay of neutral Higgs couplings have been 
studied in Ref.\cite{Ibrahim:2003jm}. In this paper we analyze 
%the SUSY QCD and
%SUSY electroweak loop corrections  
%to the couplings of the charged   Higgs with quarks and leptons in 
%MSSM. Specifically we calculate the corrections 
$\overline{\delta h_b}$,$\overline{\delta h_t}$,$\overline{\delta h_{\tau}}$,
$\overline{\Delta h_b}$,$\overline{\Delta h_t}$, and
$\overline{\Delta h_{\tau}}$ from exchange of sparticles at 
one loop. We then study their effects on the decay of 
the charged Higgs into quarks and leptons for the third family. 
We note that in the approximation 
\beqn\label{c3}
\overline{\delta h_{b,t,\tau}}= \delta h_{b,t,\tau},~~
\overline{\Delta h_{b,t,\tau}}= \Delta h_{b,t,\tau}
\eeqn
one finds  that the right hand side of  Eq.~(\ref{c2}) vanishes 
and SUSY loop correction  preserves weak isospin.
This is the approximation that is often used in the literature\cite{carena2002}..
However, in general, the equalities of Eq.~(\ref{c3}) will not
hold and there will be violations of weak isospin given by
Eq.~(\ref{c2}). In this paper we will investigate the size of
these violations and their implications on phenomena.
\begin{figure}
\hspace*{-0.6in}
\centering
\includegraphics[width=12cm,height=6cm]{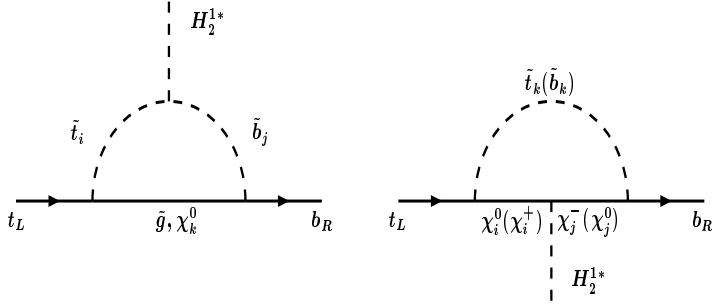}
\caption{ Exhibition of the supersymmetric loop contribution to the
 charged Higgs couplings with third generation quarks. 
 All particles in the loop are heavy supersymmetric partners with
$\tilde t_i(\tilde b_j)$ being  heavy stops (sbottoms), 
 $\tilde g$ the gluino, and $\chi_k^0(\chi_i^-)$ neutralinos (charginos).}
\label{oneloop}
\end{figure}
\section{SUSY QCD and SUSY electroweak corrections to the charged Higgs 
couplings}
Fig.~(1) gives the SUSY QCD loop correction through gluino exchange
and 
 SUSY electroweak correction via neutralino and chargino exchanges.
In the analysis of these graphs we use the zero external momentum 
approximation\cite{carena2000,carena2002,Ibrahim:2003jm,Ibrahim:2003cv}
rather than the alternate technique of 
Refs.\cite{Pierce:1996zz,Christova:2002sw}. We will work within the 
frame work of MSSM and SUGRA models allowing for nonuniversalities. 
We will characterize the parameter space of these models by using the
 parameters $m_A$, $m_0$, $\tilde m_i=m_{\frac{1}{2}}e^{i\xi_i}$ (i=1,2,3),
$A^0_t$, $A^0_b$,$A^0_{\tau}$ and $\tan\beta$. Here $m_A$ is the 
mass parameter in the Higgs sector,
 $m_0$ is the 
universal scalar mass, $\tilde m_i$ (i,2,3) are the gaugino masses 
corresponding to the gauge groups $U(1)$, $SU(2)$ and $SU(3)_C$,
$A^0_{t,b,\tau}$ are the trilinear couplings in stops, sbottoms and staus
and $\tan\beta =<H_2>/<H_1>$ where $H_2$ gives mass to the up quark and
$H_1$ gives mass to the down quark and the lepton. 
We discuss now the various contributions in detail.
For $\overline{\Delta h_b}$ we 
need $\tilde b \tilde t H$ interaction which is given
by 
\beq\label{d}
{\cal{L}}_{\tilde b \tilde t H} =H_1^2  \tilde b_j^*\tilde t_i \eta_{ji}'
+H_2^1 \tilde b_i \tilde t_j^* \eta_{ij} +H.c.
\eeq
where
\beqn\label{e}
\eta_{ji}'= \frac{gm_b}{\sqrt 2 m_W \cos\beta} m_0A_bD_{b2j}^* D_{t1i}
+\frac{gm_t}{\sqrt 2 m_W \sin\beta} \mu D_{b1j}^* D_{t2i}\nonumber\\
+\frac{gm_bm_t}{\sqrt 2 m_W \cos\beta}  D_{b2j}^* D_{t2i}
+\frac{gm_b^2}{\sqrt 2 m_W \cos\beta} D_{b1j}^* D_{t1i}
-\frac{g}{\sqrt 2} m_W \cos\beta  D_{b1j}^* D_{t1i}
\eeqn
and
\beqn\label{f}
\eta_{ij}= \frac{gm_t}{\sqrt 2 m_W \sin\beta} m_0A_tD_{b1i} D_{t2j}^*
+\frac{gm_b}{\sqrt 2 m_W \cos\beta}\mu D_{b2i} D_{t1j}^*\nonumber\\
+\frac{gm_bm_t}{\sqrt 2 m_W \sin\beta} D_{b2i} D_{t2j}^*
+\frac{gm_t^2}{\sqrt 2 m_W \sin\beta} D_{b1i} D_{t1j}^*
-\frac{g}{\sqrt 2} m_W \sin\beta  D_{b1i} D_{t1j}^*
\eeqn
Using the above one  finds from Fig.1  
\beqn\label{g}
\overline{\Delta h}_b^{\tilde g} = -\sum_{i=1}^2\sum_{j=1}^2
\frac{2\alpha_s}{3\pi} e^{-i\xi_3} D_{b2j}D_{t1i}^* \eta_{ji}^* 
m_{\tilde g} f(m_{\tilde g}^2, m_{\tilde t_i}^2,m_{\tilde b_j}^2)
\eeqn
where $f(m^2,m_i^2,m_j^2)$ is defined so that 
\beqn\label{h}
f(m^2,m_i^2,m_j^2)
= \frac {1}{(m^2-m_i^2) (m^2-m_j^2)(m_j^2-m_i^2)}\nonumber\\
(m_j^2 m^2 ln\frac{m_j^2}{m^2} 
 +m^2 m_i^2ln\frac{m^2}{m_i^2} +m_i^2 m_j^2 ln\frac{m_i^2}{m_j^2})  
 \eeqn
 for the case $i\neq j$  and 
 \beqn\label{i}
 f(m^2,m_i^2,m_i^2) =\frac {1}{(m_i^2-m^2)^2} (m^2 ln\frac{m_i^2}{m^2} 
 + (m^2-m_i^2))
\eeqn 
Further in Eq.(\ref{g}) $D_{bij}$ is the matrix that diagonalizes the 
b squark $mass^2$ matrix so that
\beqn
\tilde b_L=\sum_{i=1}^{2} D_{b1i} \tilde b_i,~~~~~ 
\tilde b_R=\sum_{i=1}^{2} D_{b2i} \tilde b_i
\eeqn
where $\tilde b_i$ are the b squark mass eigen states. 
Similarly  $D_{tij}$ is the matrix that diagonalizes the t squark 
$mass^2$ matrix so that 
\beqn
\tilde t_L=\sum_{i=1}^{2} D_{t1i} \tilde t_i,~~~~~ 
\tilde t_R=\sum_{i=1}^{2} D_{t2i} \tilde t_i
\eeqn
where $\tilde t_i$ are the t squark mass eigen states. 
The SUSY electroweak correction $\overline{\Delta h}_b^{EW}$
 arising from the neutralino exchange
 and from the chargino exchange (see Fig.1) is given by the following
\beqn\label{j}
\overline{\Delta h}_b^{EW} = \sum_{i=1}^2\sum_{j=1}^2\sum_{k=1}^4
2\eta_{ji}^* (\alpha_{bk}D_{b1j}- \gamma_{bk}D_{b2j}) 
(\beta_{tk}^* D_{t1i}^* +\alpha_{tk}D_{t2i}^*)\nonumber\\
\frac{m_{\chi_k^0}}{16\pi^2} 
f(m_{\chi_k^0}^2,m_{\tilde t_i}^2,m_{\tilde b_j}^2) 
+\sqrt 2 g\xi_{ki} \frac{m_{\chi_k^0}m_{\chi_i^-}}{16\pi^2} 
[-\kappa_b U_{i2}^*D_{t1j}
(\beta_{tk}^*D_{t1j}^* +\alpha_{tk}D_{t2j}^*)\nonumber\\
f(m_{\tilde t_j}^2,m_{\chi_i^-}^2,m_{\chi_k^0}^2) 
+ (\alpha_{bk}D_{b1j} -\gamma_{bk}D_{b2j})
 (U_{i1}^*D_{b1j}^* -\kappa_{b}U_{i2}^*D_{b2j}^*)
f(m_{\tilde b_j}^2,m_{\chi_i^-}^2,m_{\chi_k^0}^2)]
\eeqn
where
\beqn\label{k}
\xi_{ki} =-g X_{4k}V_{i1}^*- \frac{g}{\sqrt 2} X_{2k}V_{i2}^* 
-\frac{g'}{\sqrt 2} X_{1k} V_{i2}^*
\eeqn
Here U and V diagonalize the chargino mass matrix, X diagonalizes
the neutralino mass matrix, and $k_b$ is defined by
\beqn
k_{b(t)}=\frac{m_{b(t)}}{\sqrt 2 m_W\cos\beta(\sin\beta)} 
\eeqn
Finally,  $\alpha_{bk}$, $\beta_{bk}$ and $\gamma_{bk}$
for the b quark are defined so that
\beqn\label{alphabk}
\alpha_{bk} =\frac{g m_bX_{3k}}{2m_W\cos\beta},~~
\beta_{bk}=eQ_bX_{1k}^{'*} +\frac{g}{\cos\theta_W} X_{2k}^{'*}
(T_{3b}-Q_b\sin^2\theta_W)\nonumber\\
\gamma_{bk}=eQ_b X_{1k}'-\frac{gQ_b\sin^2\theta_W}{\cos\theta_W}
X_{2k}'
\eeqn
and
\beqn\label{alphatk}
\alpha_{tk} =\frac{g m_tX_{4k}}{2m_W\sin\beta},~~
\beta_{tk}=eQ_tX_{1k}^{'*} +\frac{g}{\cos\theta_W} X_{2k}^{'*}
(T_{3t}-Q_t\sin^2\theta_W)\nonumber\\
\gamma_{tk}=eQ_t X_{1k}'-\frac{gQ_t\sin^2\theta_W}{\cos\theta_W}
X_{2k}'
\eeqn
where $Q_{b(t)}= -\frac{1}{3}(\frac{2}{3})$
 and $T_{3b(t)}=-\frac{1}{2}(\frac{1}{2})$ and where
\beqn
X'_{1k}=X_{1k}\cos\theta_W +X_{2k}\sin\theta_W\nonumber\\
X'_{2k}=-X_{1k}\sin\theta_W +X_{2k}\cos\theta_W
\eeqn
Thus the total correction $\overline{\Delta h}_b$ is given by
\beqn
\overline{\Delta h}_b=  \overline{\Delta h}_b^{\tilde g}
+\overline{\Delta h}_b^{EW}
\eeqn
Similarly the SUSY QCD and SUSY electroweak correction 
$\overline{\delta h}_b$ is computed to be 
\beqn\label{l}
\overline{\delta h}_b = \sum_{i=1}^2\sum_{j=1}^2
\frac{2\alpha_s}{3\pi} e^{-i\xi_3} 
D_{b2j} D_{t1i}^* \eta_{ji}' m_{\tilde g} 
f(m_{\tilde g}^2,m_{\tilde t_i}^2,m_{\tilde b_j}^2) \nonumber\\
-\sum_{i=1}^2\sum_{j=1}^2\sum_{k=1}^4 
2 \eta_{ji}' (\alpha_{bk} D_{b1j} -\gamma_{bk} D_{b2j}) 
(\beta_{tk}^* D_{t1i}^*+ \alpha_{tk} D_{t2i}^*)
\frac{m_{\chi_k^0}}{16\pi^2} 
f(m_{\chi_k^0}^2,m_{\tilde t_i}^2,m_{\tilde b_j}^2) 
\eeqn
An analysis similar to the above gives for $\overline{\Delta h}_t$
and for $\overline{\delta h}_t$ the following
\beqn\label{m}
\overline{\Delta h}_t = -\sum_{i=1}^2\sum_{j=1}^2
\frac{2\alpha_s}{3\pi} e^{-i\xi_3} 
D_{b1i}^* D_{t2j} \eta_{ij}'^*  m_{\tilde g} 
f(m_{\tilde g}^2,m_{\tilde b_i}^2,m_{\tilde t_j}^2) \nonumber\\
+\sum_{i=1}^2\sum_{j=1}^2\sum_{k=1}^4 
2 \eta_{ij}^{'*} (\alpha_{tk} D_{t1j} -\gamma_{tk} D_{t2j}) 
(\beta_{bk}^* D_{b1i}^*+ \alpha_{bk} D_{b2i}^*)
\frac{m_{\chi_k^0}}{16\pi^2} 
f(m_{\chi_k^0}^2,m_{\tilde b_i}^2,m_{\tilde t_j}^2) 
\nonumber\\
+\sum_{i=1}^2\sum_{j=1}^2\sum_{k=1}^4 \sqrt 2 g \xi_{ki}^{*'} 
\frac{m_{\chi_k^0} m_{\chi_i^-}}{16\pi^2} 
[-k_t V_{i2}^* D_{b1j} (\beta_{bk}^*D_{b1j}^* +\alpha_{bk}D_{b2j}^*)\nonumber\\  
f(m_{\tilde b_j}^2,m_{\chi_i^-}^2,m_{\chi_k^0}^2)
+(\alpha_{tk}D_{t1j} -\gamma_{tk}D_{t2j})(V_{i1}^*D_{t1j}^* -k_tV_{i2}^*D_{t2j}^*) 
f(m_{\tilde t_j}^2,m_{\chi_i^-}^2,m_{\chi_k^0}^2)]
\eeqn
where 
\beqn\label{n}
\xi_{ki}'= -g  X_{3k}^* U_{i1} +\frac{g}{\sqrt 2} U_{i2} X_{2k}^*
+\frac{g'}{\sqrt 2} U_{i2} X_{1k}^*
\eeqn
 Similarly for $\overline{\delta h}_t$ one has the following
\beqn\label{o}
\overline{\delta h}_t = \sum_{i=1}^2\sum_{j=1}^2
\frac{2\alpha_s}{3\pi} e^{-i\xi_3} 
D_{b1i}^* D_{t2j} \eta_{ij} m_{\tilde g} 
f(m_{\tilde g}^2,m_{\tilde b_i}^2,m_{\tilde t_j}^2) \nonumber\\
-\sum_{i=1}^2\sum_{j=1}^2\sum_{k=1}^4 
2 \eta_{ij} (\alpha_{tk} D_{t1j} -\gamma_{tk} D_{t2j}) 
(\beta_{bk}^* D_{b1i}^*+ \alpha_{bk} D_{b2i}^*)
\frac{m_{\chi_k^0}}{16\pi^2} 
f(m_{\chi_k^0}^2,m_{\tilde b_i}^2,m_{\tilde t_j}^2) 
\eeqn
The analysis of $\overline{\Delta h}_{\tau}$ and of
 $\overline{\delta h}_{\tau}$ is free of the SUSY QCD correction
 while the SUSY electroweak correction gives
\beqn\label{p}
\overline{\Delta h}_{\tau} = 
\sum_{j=1}^2\sum_{k=1}^4
2 \eta_j^{\tau *} [\alpha_{\tau k} D_{\tau 1j} -\gamma_{\tau k} D_{\tau 2j}]
\beta_{\nu_k}^* \frac{m_{\chi_k^0}}{16\pi^2} 
f(m_{\chi_k^0}^2,m_{\tilde \nu}^2,m_{\tilde \tau_j}^2) \nonumber\\
- \sqrt 2 g \xi_{kj}
\frac{m_{\chi_k^0} m_{\chi_j^-}}{16\pi^2}
[k_{\tau} U_{j2}^* \beta_{\nu k}^*
f(m_{\tilde\nu}^2,m_{\chi_j^-}^2,m_{\chi_k^0}^2)]\nonumber\\
+\sum_{i=1}^2\sum_{j=1}^2\sum_{k=1}^4  \sqrt 2 g \xi_{ki} 
\frac{m_{\chi_k^0} m_{\chi_i^-}}{16\pi^2} 
[(U_{i1}^*D_{\tau1j}^*-k_{\tau}U_{i2}^* D_{\tau 2j}^*)\nonumber\\ 
(\alpha_{\tau k} D_{\tau 1j} -\gamma_{\tau k} D_{\tau 2j})
f(m_{\tilde\tau_j}^2,m_{\chi_i^-}^2,m_{\chi_k^0}^2)]
\eeqn
where $D_{\tau ij}$, $k_{\tau}$, $\alpha_{\tau k}$, $\beta_{\tau k}$, 
$\gamma_{\tau k}$, $\beta_{\nu_k}$
 are defined similar to  $D_{bij}$, $k_{b}$ etc. 
Finally, for  $\overline{\delta h}_{\tau}$ we have 
\beqn\label{q}
\overline{\delta h}_{\tau} = 
-\sum_{j=1}^2\sum_{k=1}^4
2 \eta_j^{\tau '} [\alpha_{\tau k} D_{\tau 1j} -\gamma_{\tau k} D_{\tau 2j}]
\beta_{\nu_k}^* \frac{m_{\chi_k^0}}{16\pi^2} 
f(m_{\chi_k^0}^2,m_{\tilde \nu}^2,m_{\tilde \tau_j}^2) 
\eeqn
where
\beqn\label{r}
\eta_j^{\tau} = \frac{gm_{\tau}}{\sqrt 2 m_W \cos\beta} \mu D_{\tau 2j}
-\frac{g}{\sqrt 2} m_W \sin\beta D_{\tau 1j}\nonumber\\
\eta_j^{\tau'} = \frac{gm_{\tau}}{\sqrt 2 m_W \cos\beta} m_0A_{\tau}
 D_{\tau 2j}^*
 +  \frac{gm_{\tau}^2}{\sqrt 2 m_W \cos\beta} D_{\tau 1j}^*
-\frac{g}{\sqrt 2} m_W \cos\beta D_{\tau 1j}^*
\eeqn
One measure of the size of the violation of the weak isospin is
the deviation of the barred quantities from the unbarred quantities.
Thus as  a measure of violations of weak isospin in b quark
couplings we define the quantity $r_b$ where 
\beqn
r_b= \frac{\sqrt{|\overline{\Delta h_b}|^2+|\overline{\delta h_b}|^2}}  
 {\sqrt{|\Delta h_b|^2+|\delta h_b|^2}}
\label{rb}
\eeqn
The deviation of this quantity from unity is an indication of
the violation of weak isospin in the Higgs couplings. 
Similarly we can define $r_t$ and $r_{\tau}$ by replacing
$b$ with $t$ and $\tau$ in Eq.~(\ref{rb}).

\section{SUSY loop correction to charged Higgs Decays: 
$H^-\rightarrow \bar t b$ and 
$H^-\rightarrow \bar\nu_{\tau} \tau^- $}
In this section we study the branching ratio involving the
decays $H^-\rightarrow \bar t b$ and 
$H^-\rightarrow \bar \nu_{\tau} \tau^-$.
One may recall that in the neutral Higgs sector, the ratio
$R^{h^0}=BR(h^0\rightarrow b\bar b)/BR(h^0\rightarrow \tau\bar \tau)$
is found to be sensitive to the supersymmetric loop 
corrections\cite{babu1998} and to CP phases.
In an analogous fashion in this paper we define the
ratio $R^{H^-}=BR(H^-\rightarrow  \bar t b)/ 
 BR(H^-\rightarrow  \bar \nu_{\tau}\tau^-)$ and show that this ratio
 is a sensitive function of the supersymmetric loop corrections,  
 a sensitive function of the CP phases and in addition  sensitive 
 to the violations of weak isospin. 
 To this end it is 
convenient to display the charged Higgs interaction
\beqn\label{s}
-{\cal{L}}_{int}=\bar b (B_{bt}^s+ B_{bt}^p\gamma_5)tH^-
+ \bar\tau (B_{\nu\tau}^s+ B_{\nu\tau}^p\gamma_5)\nu H^-
+ H.c.
\eeqn
where
\beqn\label{t}
B_{bt}^s =-\frac{1}{2} (h_b + \overline{\delta h_b}) e^{-i\theta_{bt}}
\sin\beta + \frac{1}{2}  \overline{\Delta h_b} e^{-i\theta_{bt}}
\cos\beta\nonumber\\
-\frac{1}{2} (h_t + \overline{\delta h_t^*}) e^{i\theta_{bt}}
\cos\beta + \frac{1}{2}  \overline{\Delta h_t^*} e^{i\theta_{bt}}
\sin\beta\nonumber\\ 
B_{bt}^p =-\frac{1}{2} (h_t + \overline{\delta h_t^*}) e^{i\theta_{bt}}
\cos\beta + \frac{1}{2}  \overline{\Delta h_t^*} e^{i\theta_{bt}}
\sin\beta\nonumber\\
+\frac{1}{2} (h_b + \overline{\delta h_b}) e^{-i\theta_{bt}}
\sin\beta - \frac{1}{2}  \overline{\Delta h_b} e^{-i\theta_{bt}}
\cos\beta\nonumber\\ 
B_{\nu\tau}^s =- B_{\nu\tau}^p =
-\frac{1}{2} (h_{\tau} + \overline{\delta h_{\tau}}) e^{-i\chi_{\tau}/2}
\sin\beta 
+ e^{-i\chi_{\tau}/2} \frac{1}{2} \overline{\Delta h_{\tau}} \cos\beta
\eeqn
where $\theta_{bt}=(\chi_b+\chi_t)/2$ and where $\chi_b,\chi_{\tau}$ and
$\chi_t$ are  defined by the following
\beqn\label{u}
\tan\chi_b =\frac{ Im (\frac{\delta h_b}{h_b} 
+ \frac{\Delta h_b}{h_b}\tan\beta)}
{1+ Re (\frac{\delta h_b}{h_b} 
+ \frac{\Delta h_b}{h_b}\tan\beta)}
\label{tanchib}
\eeqn
and the same holds for $\tan\chi_{\tau}$ with b replaced by
$\tau$ on the right hand side of Eq.~(\ref{tanchib}). 
For $\tan\chi_t$ an expression similar to Eq.~(\ref{tanchib})
holds with b replaced by  t and $\tan\beta$ replaced by 
$\cot\beta$. The coupling $h_b$ is related to the b quark mass
by the relation
\beqn\label{v}
h_b=\sqrt 2\frac{m_b}{v_1} [(1+Re \frac{\delta h_b}{h_b}
+ Re \frac{\Delta h_b}{h_b} \tan\beta)^2\nonumber\\ 
+(Im \frac{\delta h_b}{h_b}
+ Im \frac{\Delta h_b}{h_b} \tan\beta)^2]^{-\frac{1}{2}}
\label{hb}
\eeqn
%A quantity of interest is $R^{H^-}$ defined by 
%\beqn\label{w}
%R^{H^-}= \frac{\Gamma(H^-\rightarrow \bar t b)}
%{\Gamma(H^-\rightarrow \bar \nu_{\tau} \tau^-)}
%\eeqn
and similar relations hold for $h_{\tau}$.
 For $h_t$ a similar
relation holds but with $v_1$ replaced by $v_2$ and  $\tan\beta$
replaced by $\cot\beta$.
Notice that $\delta h_b$ and $\Delta h_b$ in
 Eqs.~(32, 33)
 are not barred 
quantities.
Quantities of interest for the purpose of illustration of loop effects
are  $R_{tb}$ an $R_{\nu\tau}$ defined by 
\beqn\label{rtb}
R_{tb}= \frac{\Gamma(H^-\rightarrow \bar t b)}
{\Gamma(H^-\rightarrow \bar t b)_0}
\eeqn
and
\beqn\label{rnutau}
R_{\nu\tau}=\frac{\Gamma(H^-\rightarrow \bar \nu_{\tau} \tau^-)}
{\Gamma(H^-\rightarrow \bar \nu_{\tau} \tau^-)_0}
\eeqn
where  
\beqn\label{x}
\Gamma(H^-\rightarrow \bar t b)=\frac{3}{4\pi M_{H^-}^3}
((m_t^2+m_b^2-m_{H^-}^2)^2 - 4m_t^2 m_b^2)^{\frac{1}{2}}\nonumber\\
  \{\frac{1}{2} (|B_{bt}^s|^2 + |B_{bt}^p|^2) (m_{H^-}^2-m_t^2 -m_b^2)
-\frac{1}{2} (|B_{bt}^s|^2 - |B_{bt}^p|^2)(2m_t m_b)\}(1+\omega)
\eeqn
Here $(1+\omega)$ is the QCD enhancement factor and is given 
by\cite{gorishnii}
\beqn\label{y}
(1+\omega) = 1+ 5.67 \frac{\alpha_s}{\pi} + 29.14 \frac{\alpha_s^2}{\pi^2}
\label{3g}
\eeqn
so that $(1+\omega)\simeq 1.25$ for $\alpha_S\simeq 0.12$. 
Similarly we have 
\beqn\label{z}
\Gamma(H^-\rightarrow \bar \nu_{\tau} \tau^-)=\frac{3}{8\pi M_{H^-}^3}
(m_{H^-}^2-m_{\tau}^2)^2  (|B_{\nu\tau}^s|^2 + |B_{\nu\tau}^p|^2)
\eeqn
The  quantities 
${\Gamma(H^-\rightarrow \bar t b)_0}$ and
${\Gamma(H^-\rightarrow \bar \nu_{\tau} \tau^-)_0}$
correspond to the tree value, i.e., when the loop corrections in
${\Gamma(H^-\rightarrow \bar t b)}$ and
${\Gamma(H^-\rightarrow \bar \nu_{\tau} \tau^-)}$ are set to zero.
Finally, to quantify the breakdown of the weak isospin arising 
from SUSY QCD and SUSY electroweak loop effects we define  the 
following quantities
\beqn\label{Rtbnutau}
 \Delta R_{tb/\nu\tau}= \frac{R^{H^-}-R^{H^-}_0}{R^{H^-}_0}
 \eeqn
 where the first term in the numerator includes the full loop correction
 including the effects of weak isospin violation and the quantities
 with subscripts 0 are evaluated at the tree level. Further, we define 
 \beqn\label{rtbnutau}
 \Delta r_{tb/\nu\tau}=  \frac{R^{H^-}_{nobar}-R^{H^-}_0}{R^{H^-}_0}
 \eeqn
 where $\Delta r_{tb/\nu\tau}$ is defined identical to 
 $\Delta R_{bt/\nu\tau}$ except
 that no barred quantities are  used, i.e., we set
 $\overline{\Delta h_{b,t,\tau}}= \Delta h_{b,t,\tau}$
and  $\overline{\delta h_{b,t,\tau}}= \delta h_{b,t,\tau}$ in 
Eq.~(\ref{t}).
A comparion of $\Delta r_{tb/\nu\tau}$ and $\Delta R_{bt/\nu\tau}$
exhibits the amount of weak isospin violation induced by SUSY
loop effects.
\section{Numerical analysis} 
\begin{figure}
\hspace*{-0.6in}
\centering
\includegraphics[width=8cm,height=6cm]{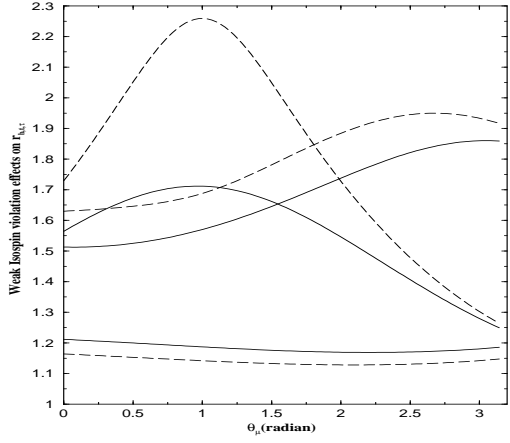}
\caption{ Plot of $r_b,r_t,r_{\tau}$ as a function of 
$\theta_{\mu}$ for the following 
inputs: (solid curves)  $m_A=200$,
 $\tan\beta =20$,  $m_0=350$, $m_{\tilde g}=300$,
 $\xi_1=.1$, $\xi_2=.2$, $\xi_3=-.3$,
 $|A^0_t|=3$, $\alpha_{A^0_t}=0$, $|A^0_b|=7$, $\alpha_{A^0_b}=2$.
 The curves in descending order at $\theta_{\mu}=0$ correspond
 to $r_b$, $r_{\tau}$ and $r_t$; (dashed curves) same input as for
 solid curves except  that $m_0=375$ and $|A^0_b|=8$. 
  All masses are in GeV and all
 angles are in GeV here and in succeeding figures.}
\label{fig1b}
\end{figure}

The analytical analysis of Secs. 2-4 is valid for MSSM.
However, the parameter space of MSSM is rather large and for
this reason for the numerical analysis we use the framework of
 SUGRA models but allowing for nonuniversalities. Specifically 
 in the numerical analysis we use the parameters 
 $m_A$, $m_0$, $A^0_t$, $A^0_b=A^0_{\tau}$ (where $A^{0'}s$ are in general
 complex), 
 $\tan\beta$ and $\xi_i$ (i=1,2,3) where $\xi_i$ are the 
phases of the gaugino masses, i.e., $\tilde m_i=m_{\frac{1}{2}}e^{i\xi_i}$.
In addition one has the Higgs mixing parameter $\mu$ (which appears in
the superpotential as $\mu H_1H_2$) which is also in general
complex, i.e., $\mu=|\mu|exp(i\theta_{\mu})$, where $ |\mu|$
is determined by radiative breaking of the electroweak symmetry
while $\theta_{\mu}$ is arbitrary.
(We note, however, that not all the phases are independent since the
phases appear only in certain combinations in physical 
quantities\cite{inmssm}). 
We then evolve them through renormalization group
equations to the low energy scales (see e.g., Ref.\cite{Arnowitt:aq}).
Now as seen from the discussion in Secs. 1 and 2, the 
weak isospin is a symmetry of the tree level Lagrangian but is violated
at the loop level. 
The size of the weak isopsin violation arising from loop corrections
can be quantified by the $r_b$ defined by 
 Eq.~(\ref{rb}) (and by $r_t, r_{\tau}$ similarly defined). 
 In Fig.~(\ref{fig1b}) we give a 
 plot of $r_b,r_t,r_{\tau}$ as a function of $\theta_{\mu}$.
 Recalling that deviations of $r_{b,t,\tau}$ 
 from unity register the violations of
 weak isospin we find that indeed such deviations can be as much
 as 50\% or more depending  on the region of the parameter
 space one is in. Thus in general the violations 
 of weak isospin arising from  Eq.(\ref{c2})
 would be significant. 
 
\begin{figure}
\hspace*{-0.6in}
\centering
\includegraphics[width=8cm,height=6cm]{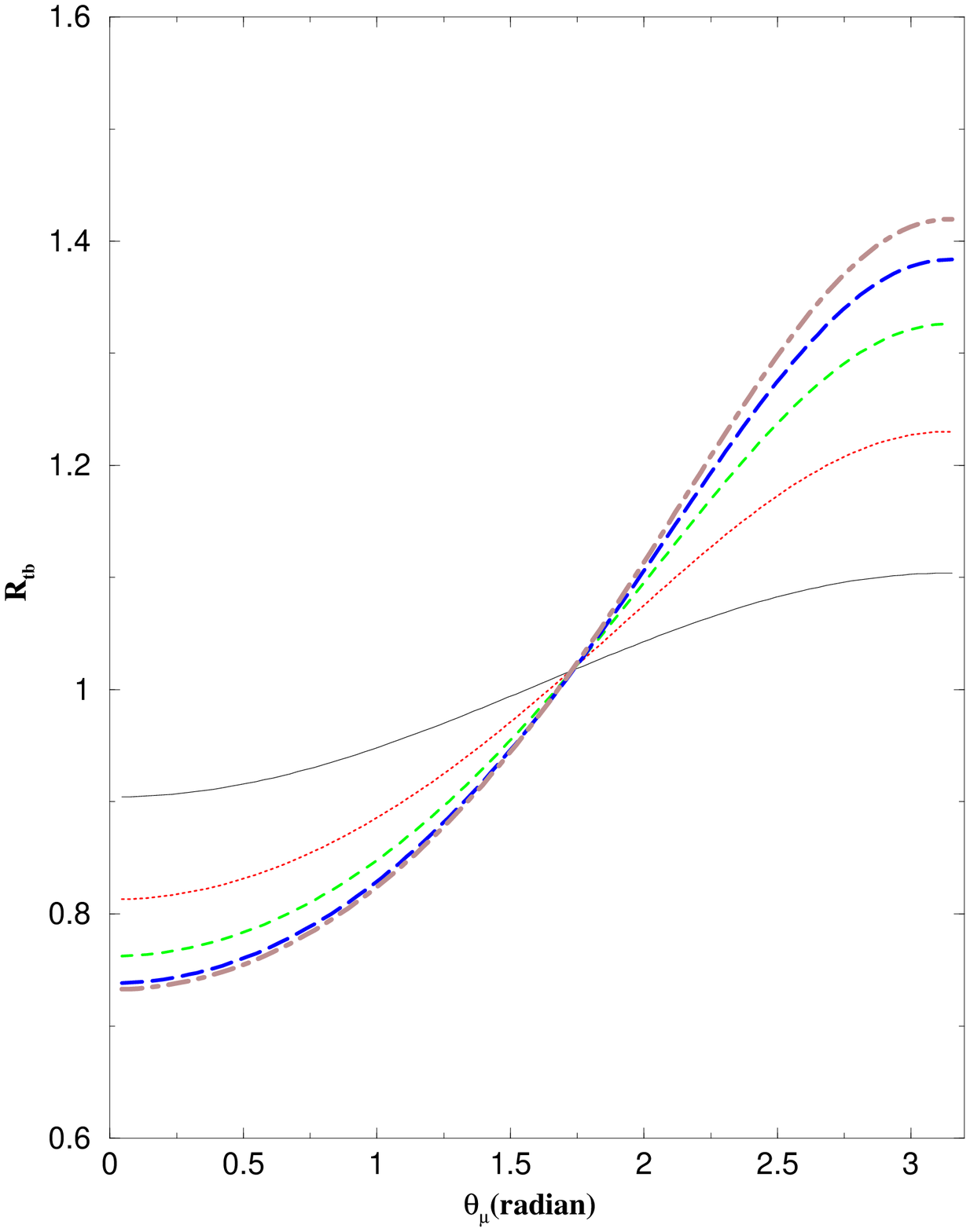}
\caption{Plot of $R_{tb}=BR(H^-\rightarrow \bar t b)_{loop}/
BR(H^-\rightarrow\bar t b)_{tree}$ 
 as a function of 
the phase $\theta_{\mu}$.  The input parameters are:  
$m_A=200$, $m_0 =200$, $m_{\tilde g}=400$, $\xi_1=0$, $\xi_2=\pi$, 
$\xi_3=\pi$, $\alpha_{A^0_t}=0= \alpha_{A^0_b}=0$, 
and $|A^0_t|= |A^0_b| =4$. The curves in ascending 
order at the point $\theta_{\mu}=\pi$ correspond to $\tan\beta$ = 
$5, 10, 15, 20, 30$.}
\label{figtb_a}
\end{figure}

\begin{figure}
\hspace*{-0.6in}
\centering
\includegraphics[width=8cm,height=6cm]{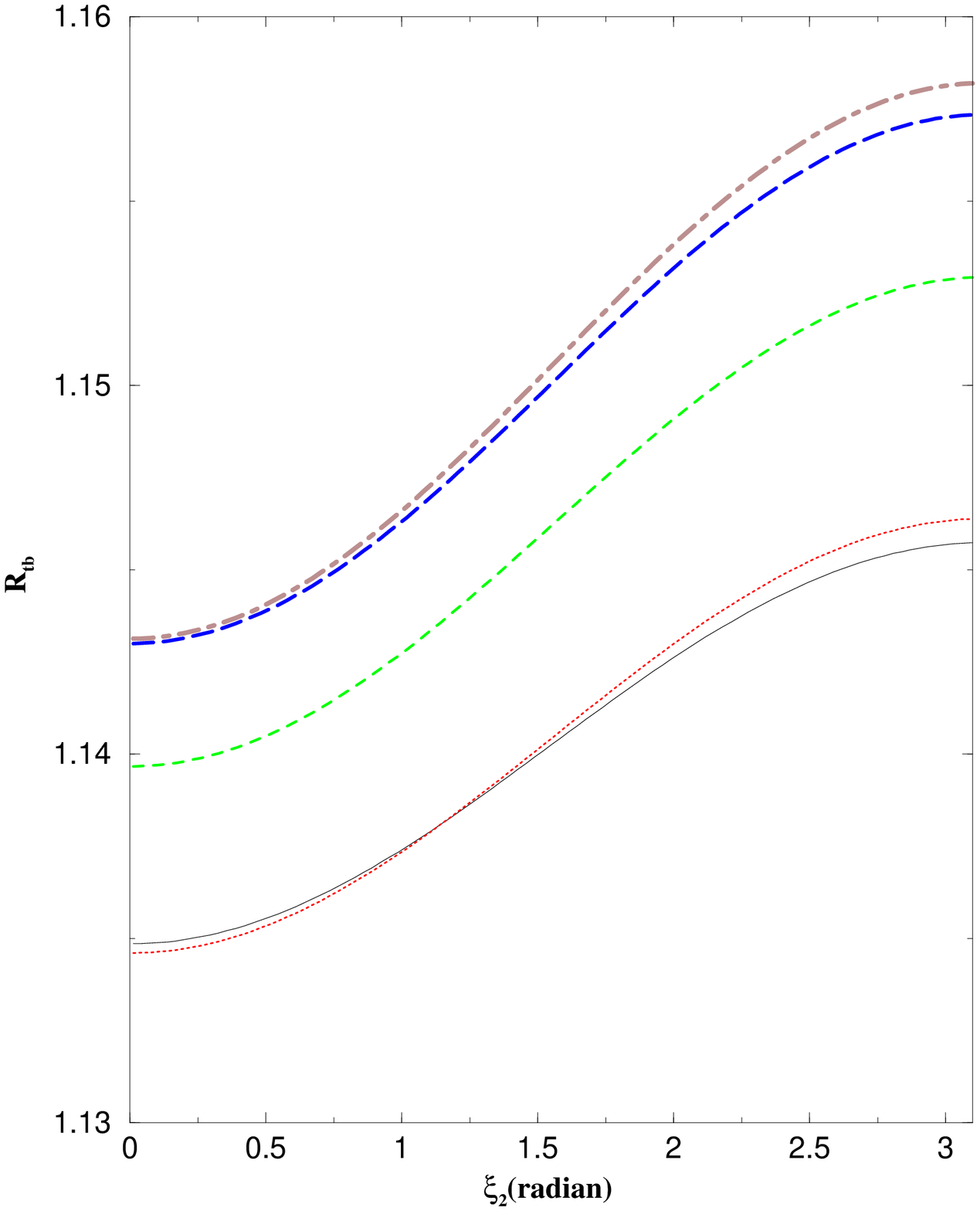}
\caption{Plot of $R_{tb}$  as a function of 
the phase $\xi_2$.  The input parameters are:  
$m_A=200$, $m_0 =200$,  $\alpha_{A^0_t}=0= \alpha_{A^0_b}=0$, 
 $|A^0_t|= |A^0_b| =2$, $\xi_1=0$, $\xi_3=0$, 
 $\theta_{\mu}=0$ and $\tan\beta =10$. The curves in ascending 
order  at  $\xi_2=\pi$ correspond to values of  
$m_{\tilde g}=300,400,600,800,1000$. }
\label{figtb_e}
\end{figure}

\begin{figure}
\hspace*{-0.6in}
\centering
\includegraphics[width=8cm,height=6cm]{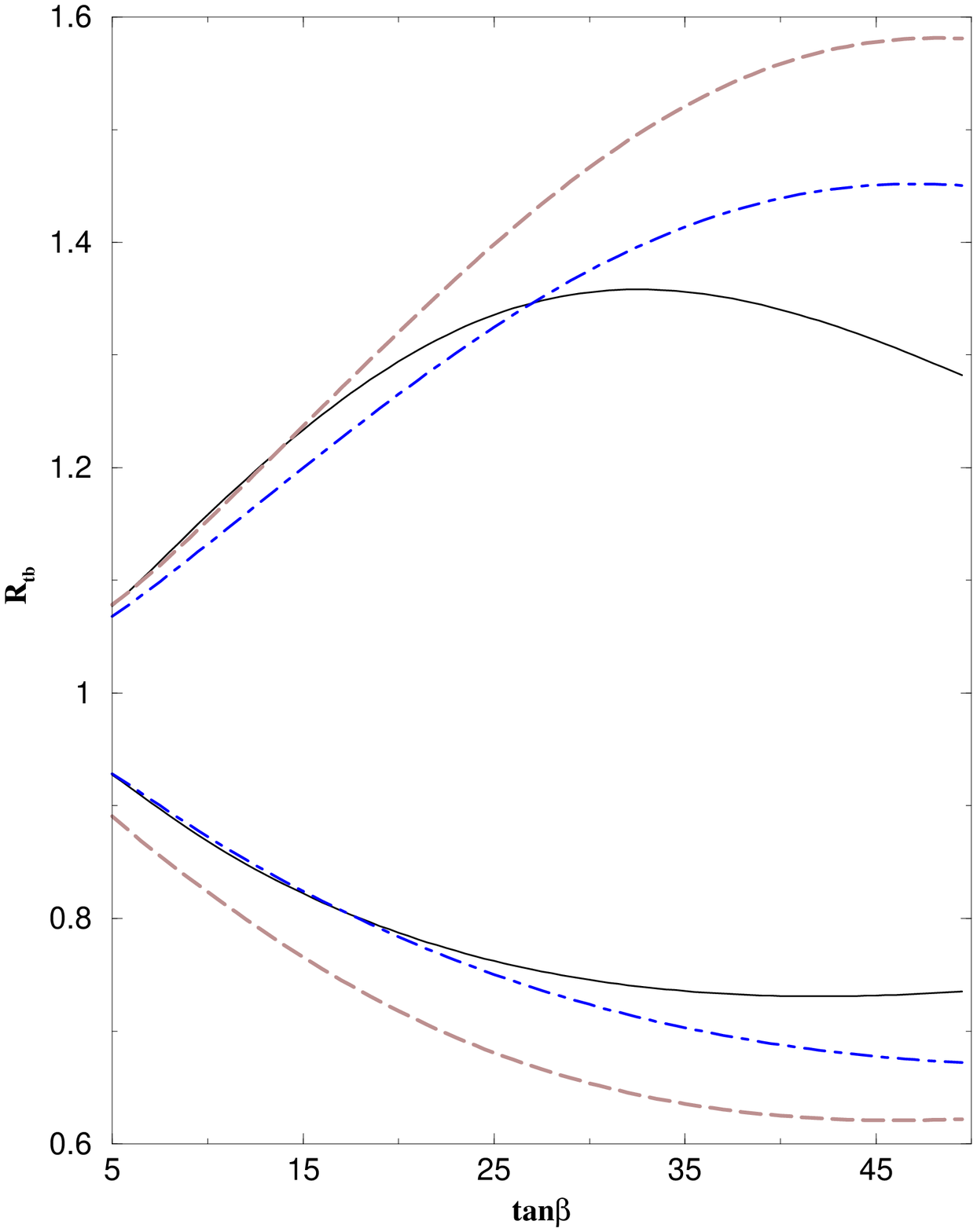}
\caption{ An exhibition of $R_{tb}$  as a function of $\tan\beta$
with and without phases. 
The  three lower curves are for the 
cases (a), (b) and (c) with parameters given by : 
(a)$m_A=200$,  $m_0=m_{\frac{1}{2}}=300$, $A_0=4$, $\alpha_{A_0}=1$,
  $\xi_1=.5$, $\xi_2=.659$, $\xi_3=.633$, $\theta_{\mu}=2.5$ (solid) ; (b) 
$m_A=200$ GeV,  $m_0=m_{\frac{1}{2}}=555$ GeV, $A_0=4$, $\alpha_{A_0}=2$,
  $\xi_1=.6$, $\xi_2=.653$, $\xi_3=.672$,$\theta_{\mu}=2.5$ (long-dashed); 
  (c) $m_A=200$ GeV,
    $m_0=m_{\frac{1}{2}}=480$ GeV, $A_0=3$, $\alpha_{A_0}=.8$,
  $\xi_1=.4$, $\xi_2=.668$, $\xi_3=.6$, $\theta_{\mu}=2.5$ (dot-dashed).
  $|A^0_t|=|A^0_b|=A_0$, $\alpha_{A^0_t}=\alpha_{A^0_b}=\alpha_{A_0}$ in 
  all cases.  The edm constraints including the $H_g^{199}$ are 
  satisfied for the above curves at $\tan\beta =50$ as shown in Table 1. 
  A plot of the above three cases but with the phases set to zero 
  are given by the similar upper curves. }
\label{figtb_f}
\end{figure}

\begin{figure}
\hspace*{-0.6in}
\centering
\includegraphics[width=8cm,height=6cm]{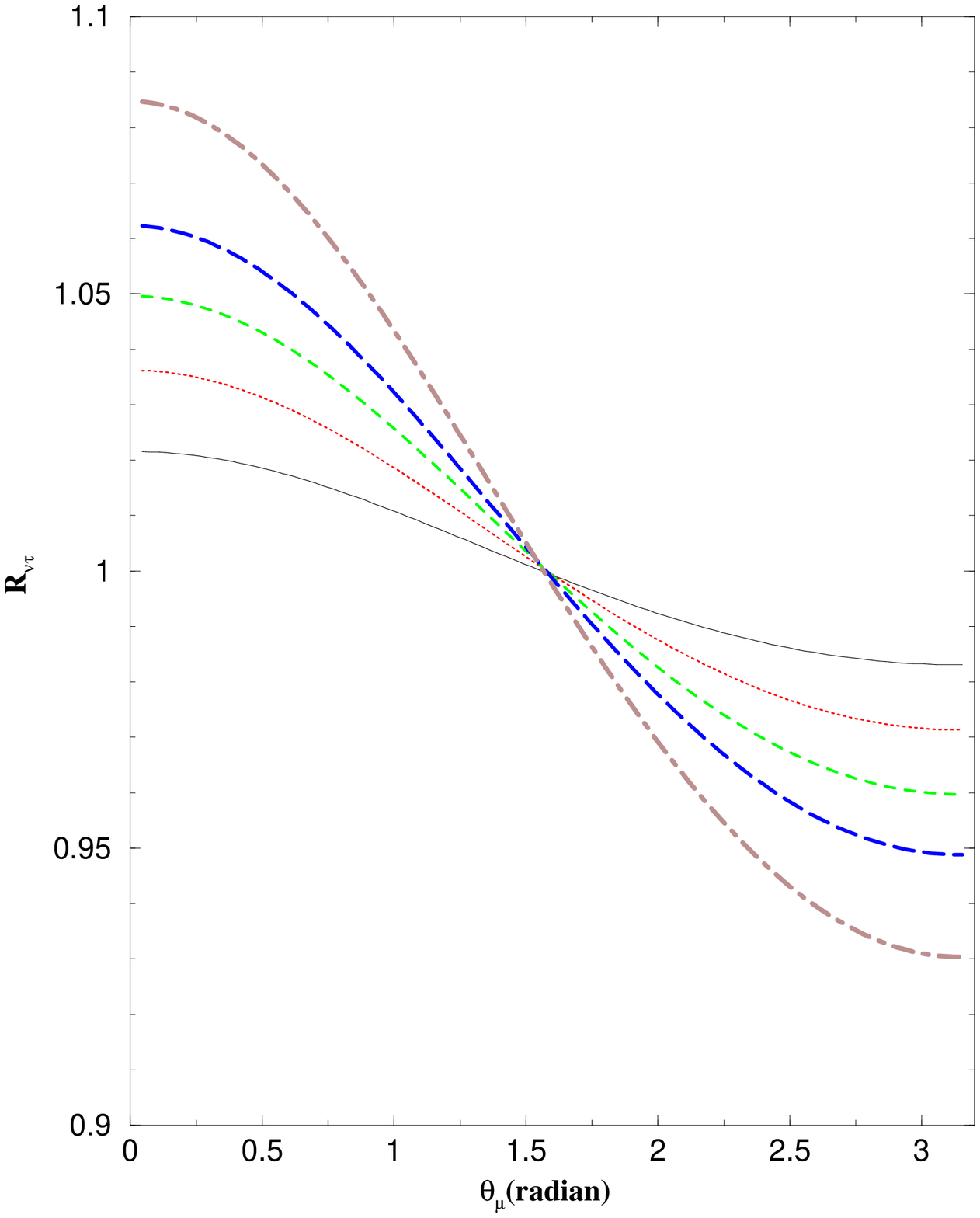}
\caption{ Plot of $R_{\nu\tau}=BR(H^-\rightarrow \bar \nu \tau^-)_{loop}/
BR(H^-\rightarrow\bar \nu \tau^-)_{tree}$ as a function of 
the phase $\theta_{\mu}$.  The input parameters are:  
$m_A=200$, $m_0 =200$, $m_{\tilde g}=400$, $\xi_1=0$, $\xi_2=\pi$, 
$\xi_3=\pi$, $\alpha_{A^0_t}= \alpha_{A^0_b}  =0$, and 
$|A^0_t|= |A^0_b|=4$. The curves in descending 
order at the point $\theta_{\mu}=\pi$ correspond to $\tan\beta$ = 
$5, 10, 15, 20, 30$ }
\label{fignutau_a}
\end{figure}

\begin{figure}
\hspace*{-0.6in}
\centering
\includegraphics[width=8cm,height=6cm]{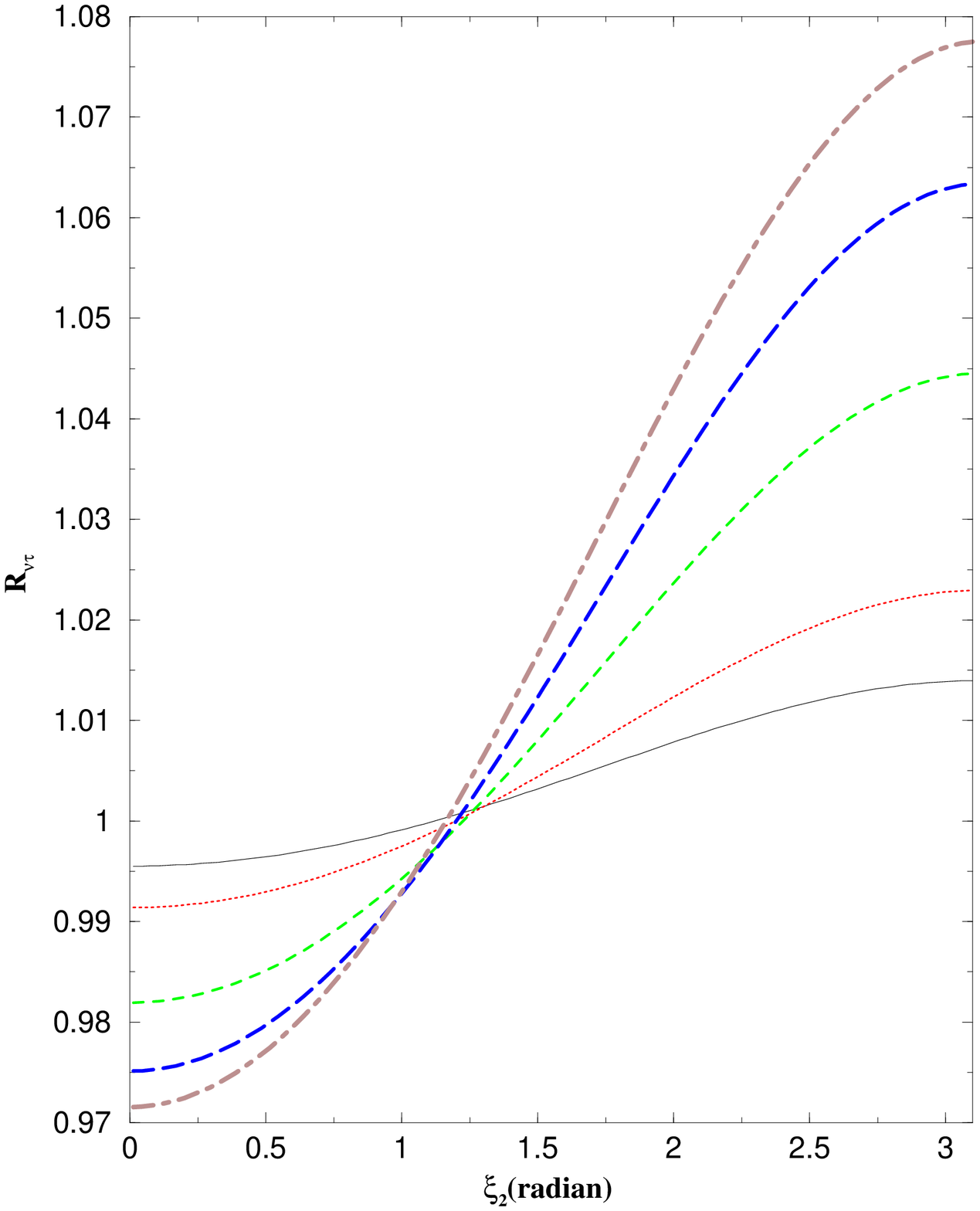}
\caption{Plot of $R_{\nu\tau}$  as a function of 
the phase $\xi_2$.  The input parameters are:  
$m_A=200$, $m_0 =200$, $\alpha_{A^0_t}= \alpha_{A^0_b}  =0$, 
$|A^0_t|= |A^0_b|=2$,
 $\xi_1=0$, $\xi_3=0$, 
 $\theta_{\mu}=0$ and $\tan\beta =10$. The curves in ascending 
order  at  $\xi_2=\pi$ correspond to values of  
$m_{\tilde g}=300,400,600,800,1000$.}
\label{fignutau_e}
\end{figure}

\begin{figure}
\hspace*{-0.6in}
\centering
\includegraphics[width=8cm,height=6cm]{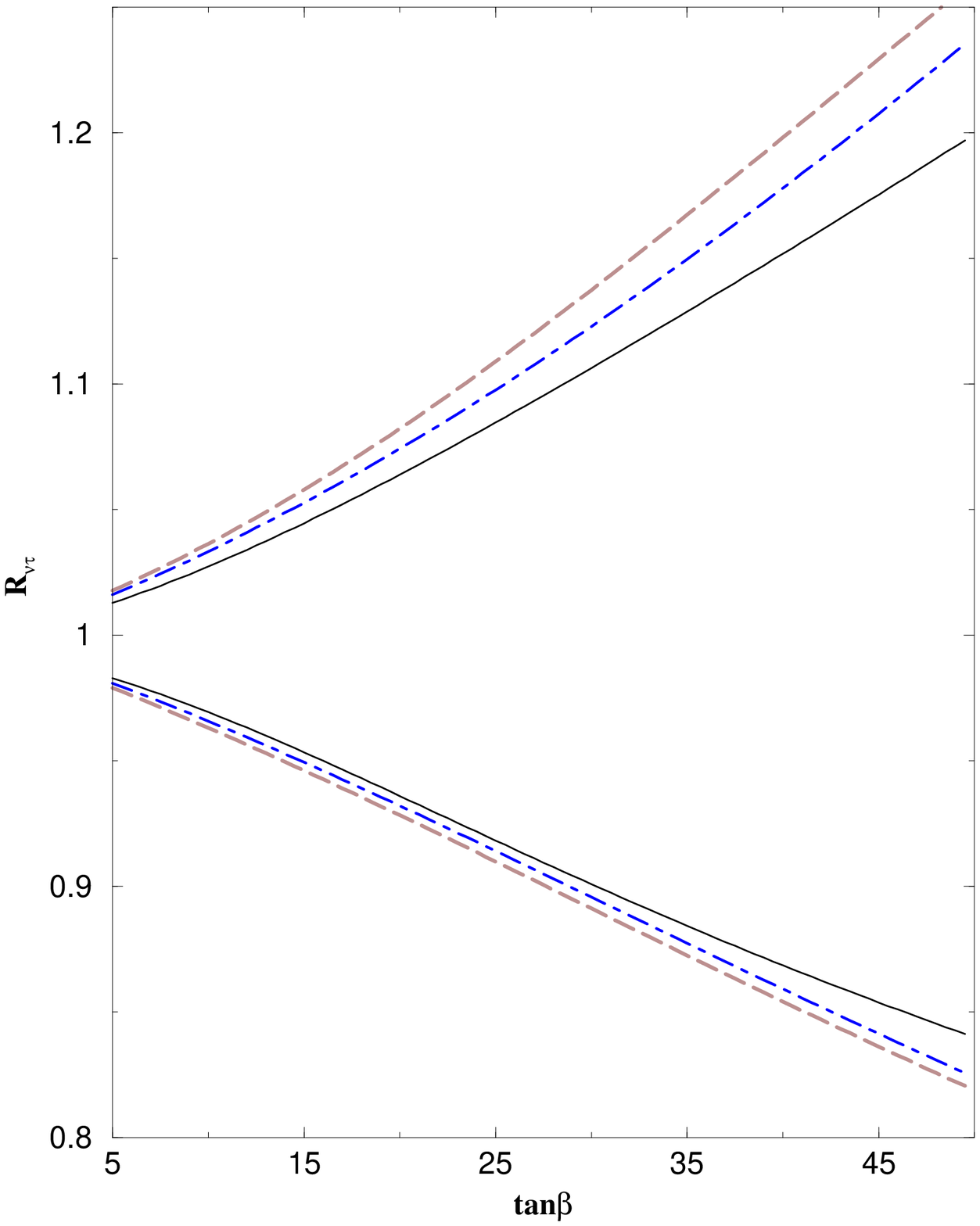}
\caption{An exhibition of $R_{\nu\tau}$  as a function of $\tan\beta$
with and without phases. 
The  three upper curves are for the 
cases (a), (b) and (c) with parameters given by : 
(a)$m_A=200$,  $m_0=m_{\frac{1}{2}}=300$, $A_0=4$, $\alpha_{A_0}=1$,
  $\xi_1=.5$, $\xi_2=.659$, $\xi_3=.633$, $\theta_{\mu}=2.5$ (solid) ; (b) 
$m_A=200$ GeV,  $m_0=m_{\frac{1}{2}}=555$ GeV, $A_0=4$, $\alpha_{A_0}=2$,
  $\xi_1=.6$, $\xi_2=.653$, $\xi_3=.672$,$\theta_{\mu}=2.5$ (long-dashed); 
  (c) $m_A=200$ GeV,
    $m_0=m_{\frac{1}{2}}=480$ GeV, $A_0=3$, $\alpha_{A_0}=.8$,
  $\xi_1=.4$, $\xi_2=.668$, $\xi_3=.6$, $\theta_{\mu}=2.5$ (dot-dashed).
 $|A^0_t|=|A^0_b|=A_0$, $\alpha_{A^0_t}=\alpha_{A^0_b}=\alpha_{A_0}$ in 
  all cases.  
    The edm constraints including the $H_g^{199}$ are 
  satisfied for the above curves at $\tan\beta =50$ as shown in Table 1. 
  A plot of the above three cases but with the phases set to zero 
  are given by the similar lower curves. }
\label{fignutau_f}
\end{figure}

\begin{figure}
\hspace*{-0.6in}
\centering
\includegraphics[width=8cm,height=6cm]{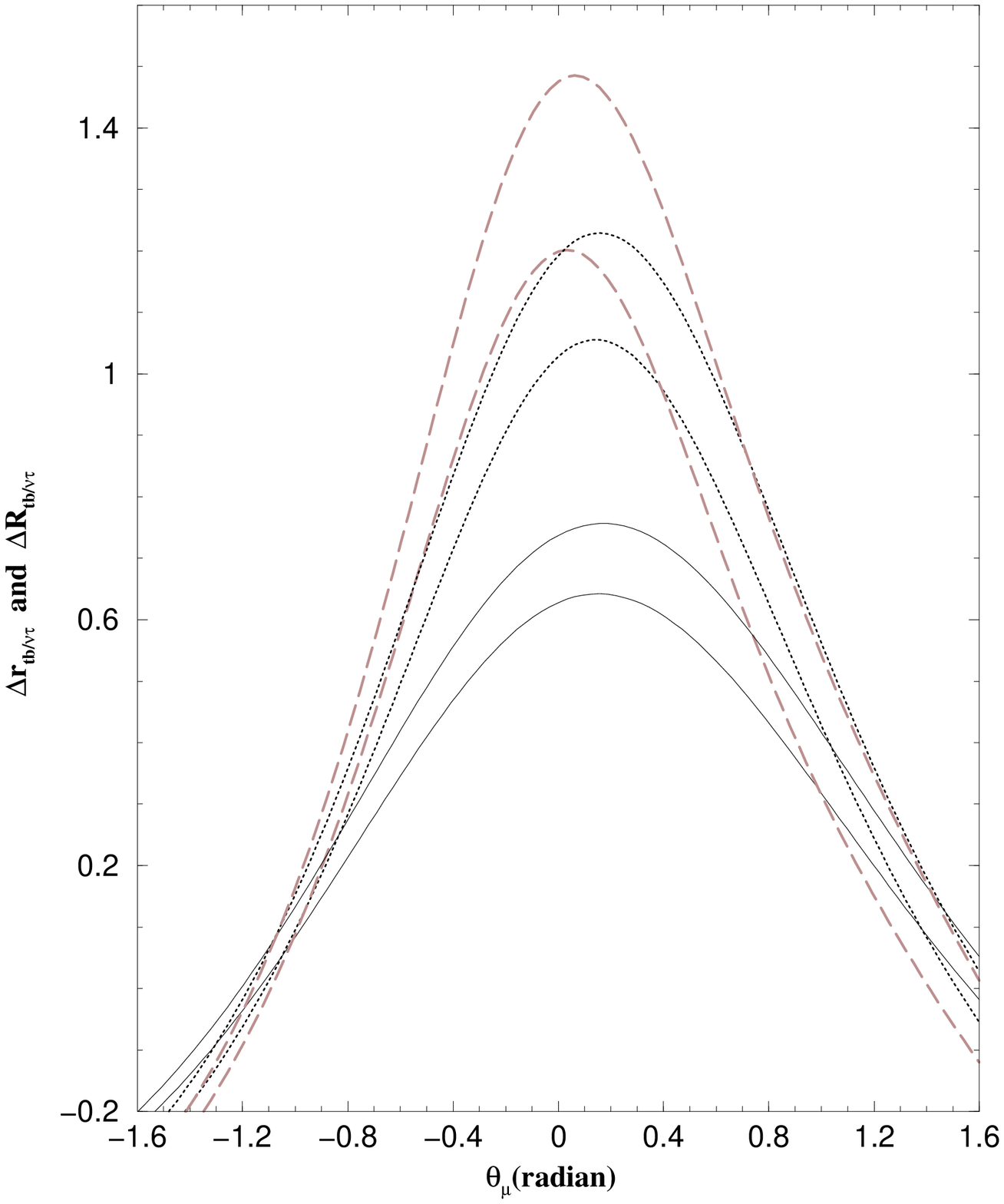}
\caption{Plot of $\Delta R_{tb/\nu\tau}$ and
$\Delta r_{tb/\nu\tau}$ as a function of
 $\theta_{\mu}$ for the following inputs:(solid curves)
$m_A=200$,
  $m_0=300$, $m_{\tilde g}=300$,
$\tan\beta =20$, $\xi_1=.1$, $\xi_2=.2$, $\xi_3=-.3$,
 $|A^0_t|=3$, $\alpha_{A^0_t}=0$, $|A^0_b|=7$, $\alpha_{A^0_b}=2$.
 The curves in descending order at $\theta_{\mu}=0$ correspond
 to $\Delta r_{tb/\nu\tau}$  and $\Delta R_{tb/\nu\tau}$;
 (dotted curves): same input as  for solid curves  except that 
 $m_0=350$; (dashed curves) same input as solid curves except
 that $m_0=375$ and $|A^0_b|=8$.}
\label{fig1cc}
\end{figure}

 Next we investigate the question of how
 large  the loop corrections themselves are relative to the
 tree values. 
In Fig.~(\ref{figtb_a}) we give a plot of $R_{tb}$ defined by
Eq.~(\ref{rtb})  as a function
of $\theta_{\mu}$ for values of $\tan\beta$ ranging from 5 to 30 for
the specific set of inputs given in caption of Fig.~(\ref{figtb_a}).
The analysis of the figure shows that the loop correction varies 
strongly with  the  phase $\theta_{\mu}$ with the  correction
changing sign as $\theta_{\mu}$  varies from 0 to $\pi$. Further, the
analysis shows that the loop correction can be as large as about 40-50\%  of 
the tree contribution in this case.
In Fig.~(\ref{figtb_e}) we  give a plot of $R_{tb}$ 
as a  function of $\xi_2$ 
%for values $m_{\tilde g}$ ranging 
%from 300 GeV to 1 TeV  
for the specific set of inputs given in the caption 
of Fig.~(\ref{figtb_e}). The analysis of Fig.~(\ref{figtb_e}) shows 
that the loop corrections are substantial and further that they have some 
sensitivity to $\xi_2$ though the sensitivity is significantly smaller 
when compared to the sensitivity to variations in $\theta_{\mu}$ 
 seen in  Fig.~(\ref{figtb_a}).
The reason for this difference is that $\xi_2$ is the phase that 
appears only in the electroweak loops whose contributions
are relatively smaller than those arising from SUSY QCD while 
the variations in $\theta_{\mu}$ arise from both QCD
and electroweak contributions.
In Fig.~(\ref{figtb_f}) we give a comparison of the loop correction to
$R_{tb}$ with and without phases as a function of $\tan\beta$. 
The lower curves are for three cases
(a), (b) and (c) whose inputs are given in the figure caption.
These cases at $\tan\beta=50$ satisfy the EDM constraints including
the $Hg^{199}$ constraint as shown in Table 1 
(taken from 
Ref.\cite{Ibrahim:2003jm}).
The upper three similar curves have all the same inputs as the lower
three curves except that the phases are all set to zero.
  Fig.~(\ref{figtb_f}) shows that the loop corrections to $R_{tb}$
  depend sensitively on the phases and the correction can 
  change sign from its tree value in the presence of phases. 
  Further, the ratio $R_{tb}$ is sensitive to
  $\tan\beta$  with or without the inclusion of phases.
The analysis of Figs  ~(\ref{fignutau_a}) - ~(\ref{fignutau_f}) 
is identical to the analysis of Figs. ~(\ref{figtb_a}) - ~(\ref{figtb_f})
except that the analysis of  Figs  ~(\ref{fignutau_a}) - ~(\ref{fignutau_f}) 
is for the ratio $R_{\nu\tau}$. 
The main difference here from the $R_{tb}$ case is that for the
case of  $R_{\nu\tau}$ there are no SUSY QCD corrections. 
Thus the sensitivity to the variations in the electroweak parameters
is larger. Thus a comparison of Fig.~(\ref{figtb_e}) and 
Fig.~(\ref{fignutau_e}) shows that $R_{\nu\tau}$ is more sensitive to
variations in $\xi_2$ than $R_{tb}$ because the electroweak 
corrections are not masked by QCD as in the case of $R_{tb}$.
Finally, in  Fig.~(\ref{fig1cc}) we plot $\Delta R_{tb/\nu\tau}$
and $\Delta r_{tb/\nu\tau}$ defined in Eq.~(\ref{Rtbnutau}) and
Eq.~(\ref{rtbnutau}) as a function of $\theta_{\mu}$. 
A comparison of the two shows that the effect of weak isospin violation
on the branching ratios can be in the neighborhood of 20-25\%.
\vspace{-.5cm}
\begin{table}[h]
\begin{center}
\caption{EDMs at $\tan\beta=50$ for Figs.(\ref{figtb_f})
and  (\ref{fignutau_f})}
\begin{tabular}{|l|l|l|l|}
\hline
Case & $|d_e| e.cm$ &  $|d_n| e.cm$ &  $C_{Hg} cm$  \\
\hline
(a)   & $1.67 \times 10^{-27}$ & $1.59 \times 10^{-27}$ & $1.18 \times 10^{-27}$  \\
\hline
(b)  & $6.05 \times 10^{-28}$ & $3.47 \times 10^{-27}$ & $1.29 \times 10^{-26}$  \\
\hline
(c) & $2.14 \times 10^{-27}$
 & $8.90 \times 10^{-28}$ & $1.25 \times 10^{-26}$\\  
\hline
\end{tabular}
\end{center}
\end{table}
\vspace{-1cm}
\section{Conclusions}
In this paper we have investigated the effects of supersymmetric
loop corrections on the violations of weak isospin in Yukawa 
couplings of the Higgs to quarks and leptons.
Specifically we have computed the gluino, chargino and neutralino 
loop corrections to the charged Higgs couplings to the third generation
quarks and leptons.
 We find  that the loop corrections to the 
charged Higgs couplings can be as much as 40-50\%  of the
tree level  contribution. We also compared the supersymmetric
loop corrections to the charged Higgs couplings with the
supersymmetric loop corrections to the neutral Higgs  couplings.
 The disparity between the charged Higgs and  the neutral Higgs
couplings is a measure of the  violations of weak isospin 
in the effective low energy Lagrangian. The analysis shows that the
effects of violations of weak isospin on the Yukawa couplings can
be as much as fifty percent or more. It is also found that such
violations are in fact also sensitive to CP phases.
Using these results we have investigated  the 
 charged Higgs decays $H^-\rightarrow  \bar t b$ and 
$H^-\rightarrow  \bar \nu_{\tau} \tau^-$. It is shown that
the branching ratios for these decays are sensitive to 
weak isospin violation effects and the effects of the violations
of weak isospin on the branching ratio can be as much as 20-25\%,
and thus accurate measurement
of the branching ratios of the charged and neutral Higgs decays
can provide a measure of such violations. 
The new results of this paper are contained in Secs.3,4 and 5.
Specifically in Sec.3 we have given computations of 
$\overline{\Delta h_{b,t,\tau}}$ and $\overline{\delta h_{b,t,\tau}}$ 
not previously computed in the literature in the current
framework. The analysis 
of this paper will also be useful in the more accurate 
computations of decays of the stops and sbottoms\cite{bartl}
and in the more accurate computation of charged Higgs 
decays\cite{cpyuan}.\\ 

\noindent
{\bf Acknowledgments}\\ 
This research was also supported in part by NSF grant PHY-0139967\\
\noindent
{\bf Appendix A} \\
For the convenience of comparison of the barred and the unbarred 
quantities we exhibit below the unbarred 
quantities for the bottom quark\cite{Ibrahim:2003jm}.
First we exhibit $\Delta h_b$. We have  
\beqn\label{b1}
\Delta h_b = - \sum_{\it i =1}^2
 \sum_{j=1}^2 \frac{2\alpha_s}{3\pi} e^{-i\xi_3}m_{\tilde g} 
G_{ij}^* D_{b1i}^{*} D_{b2j} 
f(m_{\tilde g}^2,m_{\tilde b_{\it i}}^2,m_{\tilde b_j}^2)\nonumber\\
 -\sum_{i=1}^2\sum_{j=1}^2\sum_{k=1}^2    
g^2 E_{ij}^*\{V_{k1}^*D_{t1i}^* -k_t V_{k2}^* D_{t2i}^*\}
(k_b U_{k2}^*D_{t1j})
\frac{m_{\chi_k^+}}{16\pi^2}
f(m_{\chi_k^+}^2,m_{\tilde t_i}^2,m_{\tilde t_j}^2)\nonumber\\
-\sum_{i=1}^2\sum_{j=1}^2\sum_{k=1}^2    
g^2 C_{ij}\{V_{i1}^*D_{t1k}^* -k_t V_{i2}^* D_{t2k}^*\}
(k_b U_{j2}^*D_{t1k})
\frac{m_{\chi_i^+}m_{\chi_j^+}}{16\pi^2}
f(m_{\tilde t_k}^2,m_{\chi_i^+}^2,m_{\chi_j^+}^2)\nonumber\\
+\sum_{i=1}^2\sum_{j=1}^2 \sum_{k=1}^4 
2G_{ij}^* 
\{\alpha_{bk}D_{b1j}-\gamma_{bk}D_{b2j}\} 
   \{\beta_{bk}^*D_{b1i}^*+\alpha_{bk}D_{b2i}^*\} 
  \frac{m_{\chi_k^0}}{16\pi^2}
f(m_{\chi_k^0}^2,m_{\tilde b_i}^2,m_{\tilde b_j}^2)\nonumber\\
+\sum_{i=1}^4\sum_{j=1}^4 \sum_{k=1}^2 
2\Gamma_{ij} 
\{\alpha_{bj}D_{b1k}-\gamma_{bj}D_{b2k}\} 
   \{\beta_{bi}^*D_{b1k}^*+\alpha_{bi}D_{b2k}^*\} 
  \frac{m_{\chi_i^0} m_{\chi_j^0}}{16\pi^2}
f(m_{\tilde b_k}^2, m_{\chi_i^0}^2, m_{\chi_j^0}^2 )
\eeqn
where 
\beqn
\frac{E_{ij}}{\sqrt 2} =\frac{gM_Z}{2\cos\theta_W} \{(\frac{1}{2} -\frac{2}{3}\sin^2\theta_W)
D^*_{t1i}D_{t1j} +\frac{2}{3} \sin^2\theta_W D^*_{t2i}D_{t2j}\}
\sin\beta\nonumber\\
-\frac{gm_t^2}{2M_W\sin\beta} 
[ D^*_{t1i}D_{t1j} + D^*_{t2i}D_{t2j}]
-\frac{gm_tm_0A_t}{2M_W\sin\beta} 
 D^*_{t2i}D_{t1j} 
\eeqn
and 
\beqn
\frac{C_{ij}}{\sqrt 2}= -\frac{g}{\sin\beta} 
[ \frac{m_{\chi_i^+}}{2M_W} \delta_{ij}
-Q^*_{ij} \cos\beta - R^*_{ij} ] 
\eeqn
\beqn
\frac{G_{ij}}{\sqrt 2} =\frac{gM_Z}{2\cos\theta_W} 
\{(-\frac{1}{2} +\frac{1}{3}\sin^2\theta_W)
D^*_{b1i}D_{b1j} -\frac{1}{3} \sin^2\theta_W D^*_{b2i}D_{b2j}\}
\sin\beta\nonumber\\
+\frac{gm_b\mu}{2M_W\cos\beta} D^*_{b1i}D_{b2j}
\eeqn
and where
\beqn
Q_{ij}= \sqrt{\frac{1}{2}} U_{i2}V_{j1}\nonumber\\
R_{ij}=\frac{1}{2M_W} [\tilde m_2^* U_{i1} V_{j1} + \mu^* U_{i2} V_{j2}] 
\eeqn
 $\Gamma_{ij}$ appearing in Eq.(\ref{b1}) is defined by 
\beqn
\frac{\Gamma_{ij}}{\sqrt 2}= 
-\frac{g}{2\sin\beta} [ \frac{m_{\chi_i^0}}{2M_W} \delta_{ij}
-Q^{''*}_{ij} \cos\beta
- R^{''*}_{ij}] 
\eeqn
where 
\beqn
gQ^{''}_{ij}= \frac{1}{2} [ X_{3i}^* (gX_{2j}^* -g' X_{1j}^*) +
(i\leftarrow \rightarrow j) ]\nonumber\\
R_{ij}^{''}= \frac{1}{2M_W} [ \tilde m_1^* X_{1i}^*X_{1j}^* 
+ \tilde m_2^* X_{2i}^*X_{2j}^*
-\mu^* (X_{3i}^*X_{4j}^* + X_{4i}^*X_{3j}^*) ] 
\eeqn
 Next we exhibit $\delta h_b$\cite{Ibrahim:2003jm}. We have 
\beqn
\delta h_b = - \sum_{\it i =1}^2 \sum_{j=1}^2 \frac{2\alpha_s}{3\pi} e^{-i\xi_3}m_{\tilde g} 
H_{ji} D_{b1i}^{*} D_{b2j} 
f(m_{\tilde g}^2,m_{\tilde b_{\it i}}^2,m_{\tilde b_j}^2)\nonumber\\
 -\sum_{i=1}^2\sum_{j=1}^2\sum_{k=1}^2    
g^2 F_{ji}\{V_{k1}^*D_{t1i}^* -k_t V_{k2}^* D_{t2i}^*\}
(k_b U_{k2}^*D_{t1j})
\frac{m_{\chi_k^+}}{16\pi^2}
f(m_{\chi_k^+}^2,m_{\tilde t_i}^2,m_{\tilde t_j}^2)\nonumber\\
+\sum_{i=1}^2\sum_{j=1}^2 \sum_{k=1}^4 
2H_{ji} 
\{\alpha_{bk}D_{b1j}-\gamma_{bk}D_{b2j}\} 
   \{\beta_{bk}^*D_{b1i}^*+\alpha_{bk}D_{b2i}^*\} 
  \frac{m_{\chi_k^0}}{16\pi^2}
f(m_{\chi_k^0}^2,m_{\tilde b_i}^2,m_{\tilde b_j}^2)
\eeqn
where 
\beqn
\frac{H_{ij}}{\sqrt 2} =-\frac{gM_Z}{2\cos\theta_W} 
\{(-\frac{1}{2} +\frac{1}{3}\sin^2\theta_W)
D^*_{b1i}D_{b1j} -\frac{1}{3} \sin^2\theta_W D^*_{b2i}D_{b2j}\}
\cos\beta\nonumber\\
-\frac{gm_b^2}{2M_W\cos\beta} 
[ D^*_{b1i}D_{b1j} + D^*_{b2i}D_{b2j}]
-\frac{gm_bm_0A_b}{2M_W\cos\beta} 
 D^*_{b2i}D_{b1j} 
\eeqn
and
\beqn
\frac{F_{ij}}{\sqrt 2} =-\frac{gM_Z}{2\cos\theta_W}
 \{(\frac{1}{2} -\frac{2}{3}\sin^2\theta_W)
D^*_{t1i}D_{t1j} +\frac{2}{3} \sin^2\theta_W D^*_{t2i}D_{t2j}\}
\cos\beta\nonumber\\
+\frac{gm_t\mu}{2M_W\sin\beta} 
 D^*_{t1i}D_{t2j}
\eeqn

\newpage

%\newpage

%\newpage

%%%%%%%%%

%\newpage

\begin{thebibliography}{999}
\bibitem{carena2002}
For a recent review, see, \\
M.~Carena and H.~E.~Haber,
%``Higgs boson theory and phenomenology. ((V)),''
Prog.\ Part.\ Nucl.\ Phys.\  {\bf 50}, 63 (2003)
[arXiv:hep-ph/0208209].

\bibitem{eedm}
E. Commins, et. al., Phys. Rev. {\bf A50}, 2960(1994).

\bibitem{nedm}
P.G. Harris et.al., Phys. Rev. Lett. {\bf 82}, 904(1999).

\bibitem{atomic}
S.~K.~Lamoreaux, J.~P.~Jacobs, B.~R.~Heckel, F.~J.~Raab and E.~N.~Fortson,
%``New Limits On Spatial Anisotrophy From Optically Pumped He-201 And Hg-199,''
Phys.\ Rev.\ Lett.\  {\bf 57}, 3125 (1986).
%%CITATION = PRLTA,57,3125;%%

 \bibitem{incancel}
T. Ibrahim and P. Nath,  Phys.\ Lett.\ B {\bf 418}, 98 (1998); 
 Phys. Rev. {\bf D57}, 478(1998); Phys. Rev. {\bf D58}, 111301(1998);
 T. Falk and K Olive, Phys. Lett. {\bf B 439}, 71(1998);
 M. Brhlik, G.J. Good, and G.L. Kane, Phys. Rev. {\bf D59}, 115004
 (1999); A. Bartl, T. Gajdosik, W. Porod, P. Stockinger, and
 H. Stremnitzer,  Phys. Rev. {\bf 60}, 073003(1999);
 S. Pokorski, J. Rosiek and C.A. Savoy, 
 Nucl.Phys. {\bf B570}, 81(2000);
 E.~Accomando, R.~Arnowitt and B.~Dutta,
%``Grand unification scale CP violating phases and the electric dipole  moment,''
Phys.\ Rev.\ D {\bf 61}, 115003 (2000);
%[arXiv:hep-ph/9907446].
  U. Chattopadhyay, T. Ibrahim, D.P. Roy, Phys.Rev.D64:013004,2001;
 C.~S.~Huang and W.~Liao,
%``Supersymmetric CP violation in B $\to$ X/s l+ l- in
% minimal supergravity  model,''
Phys.\ Rev.\ D {\bf 61}, 116002 (2000);
%%CITATION = HEP-PH 9908246;%%
ibid, Phys.\ Rev.\ D {\bf 62}, 016008 (2000);
 A.Bartl, T. Gajdosik, E.Lunghi, A. Masiero, W. Porod,
H. Stremnitzer and O. Vives, hep-ph/0103324.
%~ M. Graesser and 
%S. Thomas, hep-ph/0104254. 
% For analyses in the context string and brane models see,
 M. Brhlik, L. Everett, G. Kane and J. Lykken, Phys. Rev.
 Lett. {\bf 83}, 2124, 1999; Phys. Rev. {\bf D62}, 035005(2000);
  E. Accomando, R. Arnowitt and B. Datta, 
Phys. Rev. {\bf D61},  075010(2000);
T. Ibrahim and P. Nath, Phys. Rev. {\bf D61}, 093004(2000).

\bibitem{olive} 
 T. Falk, K.A. Olive, M. Prospelov, and R. Roiban, Nucl. Phys. 
 {\bf B560}, 3(1999); V.~D.~Barger, T.~Falk, T.~Han, J.~Jiang, T.~Li 
 and T.~Plehn,
%``CP-violating phases in SUSY, electric dipole moments, and linear  colliders,''
Phys.\ Rev.\ D {\bf 64}, 056007 (2001);
%[arXiv:hep-ph/0101106].
S.Abel, S. Khalil, O.Lebedev, Phys. Rev. Lett. {\bf 86}, 5850(2001);
T.~Ibrahim and P.~Nath,
%``CP violation effects on B/(s,d)0 $\to$ l+ l- in supersymmetry at large 
% tan(beta),''
Phys.\ Rev.\ D {\bf 67}, 016005 (2003)
.

\bibitem{na} 
P. Nath, Phys. Rev. Lett.{\bf 66}, 2565(1991); 
Y. Kizukuri and  N. Oshimo, Phys.Rev.{\bf D46},3025(1992).

\bibitem{chang}
D. Chang, W-Y.Keung,and A. Pilaftsis, Phys. Rev. Lett. {\bf 82}, 
900(1999). 

\bibitem{Ibrahim:2002ry}
For a more  complete set of references see, 
T.~Ibrahim and P.~Nath,
``Phases and CP violation in SUSY,''
arXiv:hep-ph/0210251 published in 
P.~Nath and P.~M.~.~Zerwas,
``Supersymmetry and unification of fundamental interactions. 
Proceedings, 10th International Conference, SUSY'02, Hamburg, Germany, June 17-23,
2002,'' DESY-PROC-2002-02

\bibitem{yamaguchi}
For a recent analysis for the generation of sizalble phases in a specific 
supergravity scenario see, 
M.~Endo, M.~Kakizaki and M.~Yamaguchi,
%``Radiative CP phases in Supergravity Theories,''
arXiv:hep-ph/0311206.


\bibitem{pilaftsis}
A. Pilaftsis, Phys. Rev. {\bf D58}, 096010; Phys. Lett.{\bf B435}, 
88(1998);
~A. Pilaftsis and C.E.M. Wagner, Nucl. Phys. {\bf B553}, 3(1999);
~D.A. Demir, Phys. Rev. {\bf D60}, 055006(1999);
~S.~Y.~Choi, M.~Drees and J.~S.~Lee,
%``Loop corrections to the neutral Higgs boson sector of the MSSM 
%with  explicit CP violation,''
Phys.\ Lett.\ B {\bf 481}, 57 (2000);
%[arXiv:hep-ph/0002287];
%%CITATION = HEP-PH 0002287;%%
~M.~Boz,
%``The Higgs masses and explicit CP violation in the gluino axion model,''
Mod.\ Phys.\ Lett.\ A {\bf 17}, 215 (2002).
%[arXiv:hep-ph/0008052].
%%CITATION = HEP-PH 0008052;%%

\bibitem{inhiggs}
T. Ibrahim and P. Nath,  
Phys.Rev.D63:035009,2001; hep-ph/0008237; 
T.~Ibrahim,
%``Mixing of the CP even and the CP odd Higgs bosons and the EDM  constraints,''
Phys.\ Rev.\ D {\bf 64}, 035009 (2001);
%[arXiv:hep-ph/0102218];
T.~Ibrahim and P.~Nath,
%``Neutralino exchange corrections to the Higgs boson mixings with  explicit CP violation,''
Phys.\ Rev.\ D {\bf 66}, 015005 (2002);
%[arXiv:hep-ph/0204092].
%%CITATION = HEP-PH 0204092;%%
~S.~W.~Ham, S.~K.~Oh, E.~J.~Yoo, C.~M.~Kim and D.~Son,
%``The one-loop correction to the neutral Higgs boson of the minimal
%  supersymmetric standard model in explicit CP violation scenario,''
arXiv:hep-ph/0205244.
%%CITATION = HEP-PH 0205244;%%

\bibitem{Carena:2001fw}
M.~Carena, J.~R.~Ellis, A.~Pilaftsis and C.~E.~Wagner,
%``Higgs-boson pole masses in the MSSM with explicit CP violation,''
Nucl.\ Phys.\ B {\bf 625}, 345 (2002)
[arXiv:hep-ph/0111245].
%%CITATION = HEP-PH 0111245;%%
;
M.~Carena, J.~R.~Ellis, S.~Mrenna, A.~Pilaftsis and C.~E.~M.~Wagner,
%``Collider probes of the MSSM Higgs sector with explicit CP violation,''
Nucl.\ Phys.\ B {\bf 659}, 145 (2003)
[arXiv:hep-ph/0211467].

\bibitem{msugra}
A.H. Chamseddine, R. Arnowitt and P. Nath, \Journal{\PRL}{49}
{970}{1982}; ~R. Barbieri, S. Ferrara and C.A. Savoy, \Journal{\PLB}
{119}{343}{1982}; ~L. Hall, J. Lykken, and S. Weinberg,
\Journal{\PRD}{27}{2359}{1983}:~ P. Nath, R. Arnowitt and A.H. Chamseddine,
\Journal{\NPB}{227}{121}{1983}.
% For reviews, see P. Nath, R. Arnowitt
%and A.H. Chamseddine, "Applied N=1 Supergravity", world scientific,
%1984; H.P. Nilles, Phys. Rep. {\bf 110}, 1(1984).

\bibitem{gunion}
J.~F.~Gunion and H.~E.~Haber,
%``Higgs Bosons In Supersymmetric Models. 1,''
Nucl.\ Phys.\ B {\bf 272}, 1 (1986)
[Erratum-ibid.\ B {\bf 402}, 567 (1993)].
%%CITATION = NUPHA,B272,1;%%

\bibitem{Ibrahim:2003ca}
T.~Ibrahim and P.~Nath,
%``Supersymmetric QCD and supersymmetric electroweak loop corrections to b, t, and tau masses including the effects of CP phases,''
Phys.\ Rev.\ D {\bf 67}, 095003 (2003)
[Erratum-ibid.\ D {\bf 68}, 019901 (2003)]
[arXiv:hep-ph/0301110].
%%CITATION = HEP-PH 0301110;%%

\bibitem{Ibrahim:2003cv}
T.~Ibrahim and P.~Nath,
%``Corrections to b, t quark masses and tau lepton mass in SUGRA including CP phases,''
arXiv:hep-ph/0308167.
%

\bibitem{Ibrahim:2003jm}
T.~Ibrahim and P.~Nath,
%``Decays of Higgs to b anti-b, tau anti-tau and c anti-c as signatures of  supersymmetry and CP phases,''
Phys.\ Rev.\ D {\bf 68}, 015008 (2003)
[arXiv:hep-ph/0305201].
%%CITATION = HEP-PH 0305201;%%

\bibitem{carena2000}
M.~Carena, D.~Garcia, U.~Nierste and C.~E.~Wagner,
%``Effective Lagrangian for the anti-t b H+ interaction in the MSSM and  charged Higgs phenomenology,''
Nucl.\ Phys.\ B {\bf 577}, 88 (2000)
[arXiv:hep-ph/9912516].

\bibitem{Pierce:1996zz}
D.~M.~Pierce, J.~A.~Bagger, K.~T.~Matchev and R.~j.~Zhang,
%``Precision corrections in the minimal supersymmetric standard model,''
Nucl.\ Phys.\ B {\bf 491}, 3 (1997)
[arXiv:hep-ph/9606211].

\bibitem{Christova:2002sw}
E.~Christova, H.~Eberl, W.~Majerotto and S.~Kraml,
%``CP violation in charged Higgs boson decays into tau and neutrino,''
JHEP {\bf 0212}, 021 (2002)
[arXiv:hep-ph/0211063];
E.~Christova, H.~Eberl, W.~Majerotto and S.~Kraml,
%``CP violation in charged Higgs decays in the MSSM with complex  parameters,''
Nucl.\ Phys.\ B {\bf 639}, 263 (2002)
[Erratum-ibid.\ B {\bf 647}, 359 (2002)]
[arXiv:hep-ph/0205227].

\bibitem{babu1998}
K.~S.~Babu and C.~F.~Kolda,
%``Signatures of supersymmetry and Yukawa unification in Higgs decays,''
Phys.\ Lett.\ B {\bf 451}, 77 (1999);
M.~Carena, S.~Mrenna and C.~E.~Wagner,
%``MSSM Higgs boson phenomenology at the Tevatron collider,''
Phys.\ Rev.\ D {\bf 60}, 075010 (1999).

\bibitem{gorishnii}
S.~G.~Gorishnii, A.~L.~Kataev, S.~A.~Larin and L.~R.~Surguladze,
%``Corrected Three Loop QCD Correction To The Correlator Of The Quark Scalar Currents And Gamma (Tot) (H0 $\to$ Hadrons),''
Mod.\ Phys.\ Lett.\ A {\bf 5}, 2703 (1990).
%%CITATION = MPLAE,A5,2703;%%


 \bibitem{inmssm}
T. Ibrahim and P. Nath,  Phys. Rev. {\bf D58}, 111301(1998).

\bibitem{Arnowitt:aq}
R.~Arnowitt and P.~Nath,
%``Susy Mass Spectrum In SU(5) Supergravity Grand Unification,''
Phys.\ Rev.\ Lett.\  {\bf 69}, 725 (1992).
%%CITATION = PRLTA,69,725;%%

\bibitem{bartl}
A.~Bartl, S.~Hesselbach, K.~Hidaka, T.~Kernreiter and W.~Porod,
%``Impact of SUSY CP phases on stop and sbottom decays in the MSSM,''
arXiv:hep-ph/0306281. 

\bibitem{cpyuan}
Q.~H.~Cao, S.~Kanemura and C.~P.~Yuan,
%``Associated Production of CP-odd and Charged Higgs Bosons at Hadron Colliders,''
arXiv:hep-ph/0311083.

\end{thebibliography}
\end{document}